\documentclass{midl} % Include author names
\usepackage{mwe} % to get dummy images
\usepackage{float} %Rune
\usepackage{xcolor} %Rune

\jmlrproceedings{MIDL}{Medical Imaging with Deep Learning}
\jmlrpages{}
\jmlryear{2019}
\jmlrworkshop{MIDL 2019 -- Extended Abstract Track}

\title[Multiscale Tissue Classification]{Multiscale Deep Neural Networks for Multiclass Tissue Classification of Histological Whole-Slide Images}

\midlauthor{\Name{Rune Wetteland\nametag{$^{1}$}} \Email{rune.wetteland@uis.no}\\
\Name{Kjersti Engan\nametag{$^{1}$}} \Email{kjersti.engan@uis.no}\\
\Name{Trygve Eftest\o l\nametag{$^{1}$}} \Email{trygve.eftestol@uis.no}\\
\Name{Vebj\o rn Kvikstad\nametag{$^{2}$}} \Email{vebjorn.kvikstad@sus.no}\\
\Name{Emilius {A.M. Janssen}\nametag{$^{2,3}$}} \Email{emilius.adrianus.maria.janssen@sus.no}\\
\addr $^{1}$Department of Electrical Engineering and Computer Science, University of Stavanger, Norway \\
\addr $^{2}$Department of Pathology, Stavanger University Hospital, Norway \\
\addr $^{3}$Department of Mathematics and Natural Sciences, University of Stavanger, Norway
}

\begin{document}

\maketitle

\begin{abstract}
Correct treatment of urothelial carcinoma patients is dependent on accurate grading and staging of the cancer tumour. This is determined manually by a pathologist by examining the histological whole-slide images (WSI). The large size of these images makes this a time-consuming and challenging task. The WSI contain a variety of tissue types, and a method for defining diagnostic relevant regions would have several advantages for visualization as well as further input to automated diagnosis systems. We propose an automatic multiscale method for classification of tiles from WSI of urothelial carcinoma patients into six classes. Three architectures based on convolutional neural network (CNN) were tested: MONO-CNN (400x), DI-CNN (100x/400x) and TRI-CNN (25x/100x/400x). The preliminary results show that the two multiscale models performed significantly better than the mono-scale model, achieving an F1-score of 0.986, substantiating that utilising multiple scales in the model aids the classification accuracy.
\end{abstract}

\section{Introduction}
Bladder cancer is the 10th most common cancer type worldwide \cite{GLOBOCAN2018}. More than 90\% of bladder cancer cases are urothelial carcinomas which has a particular high recurrence (50-70\%) and progression rate (10-30\%), making correct treatment and follow-up vital for survivability. Treatment is dependent on the cancer grade and stage, determined manually by an expert pathologist examining the histological whole-slide images (WSI). This is a time-consuming and challenging task, and studies have shown that it may have a low reproducibility in some cases, such as grading of urothelial carcinoma \cite{Mangrud2014}.

Examination of the WSI is challenging because of the large size of the image, which contains several different tissue types, where only some are useful for diagnostic information. An automatic tool for identification of such regions would be beneficial for both guiding a pathologist to the useful areas of the large WSI during examination, and for ROI extraction of useful tissue for a computer aided diagnostic solution. In this paper we present an automatic method for classification of tiles from WSI of urothelial carcinoma patients into the classes: urothelium, stroma, muscle, damaged tissue, blood and background. The tiles are extracted at different magnification levels, to combine and utilise information at different scales in a similar fashion to that of a pathologist. 

Multiscale approaches to tile-based classification have previously been done on other cancer types. In the work of \citet{Li2017} a multiscale U-Net was proposed for segmentation of histological images from radical prostatectomies to classify tiles into four classes. Tiles of size 100x100, 200x200 and 400x400 pixels were all extracted from histological images at 200x magnification. Features from the different tiles were then concatenated and used as input to the multiscale U-net. The model achieved a mean Jaccard index of 65.8\% over the four classes. In \citet{sirinukunwattana2018improving} a comparison of five single-scale and five multiscale architectures were tested on two datasets. Their best model (G) was a multiscale model which achieved an average F1-score of 0.782$\pm$0.07 across four classes of prostate cancer and 0.538$\pm$0.08 across four classes of breast cancer. Their result supports the claim that incorporating a larger visual context improves the results. In \citet{wetteland2019} we presented a method based on deep convolutional neural networks (CNN) for classifying tiles of urothelial carcinoma WSI into the six classes mentioned above. This was a single-scale approach where all tiles were extracted from the full resolution image of 400x magnification. The method got an F1-score of 0.934$\pm$0.061.

\section{Data Material}
The data material consists of Hematoxylin Eosin Saffron (HES) stained WSI from patients diagnosed with primary papillary urothelial carcinoma, collected at the University Hospital of Stavanger, Norway. An expert pathologist has carefully annotated 239 selected regions from 50 WSI from 32 unique patients, where each region includes one of the five foreground classes. Regions belonging to the background class was annotated on seven randomly selected patients.

Tiles were extracted from these regions at 25x, 100x and 400x magnification in such a manner that the centre pixel is the same in all three tiles. All tiles have the same size of 128x128x3 pixels. Tiles belonging to the test set was extracted from patients not present in the training data. The remaining data was augmented to balance the dataset and was further randomly shuffled and split into 85\% training and 15\% validation data. A random seed was set to ensure that the shuffling was the same for each model. The final datasets consist of 128K training tiles, 23K validation tiles and 11K test tiles. 

\section{Method and Results}
This paper compares three architectures referred to as the MONO-, DI- and TRI-CNN model. The three architectures have one (400x), two (100x, 400x) and three (25x, 100x, 400x) inputs, respectively. Each input is fed into a pre-trained VGG16 network \cite{Simonyan2014} which acts as a feature extractor. The fully-connected (FC) layers of VGG16 are replaced with a classification network consisting of two FC-layers, each followed by a dropout layer, and a final softmax layer with one output node for each of the six classes. The DI-CNN and TRI-CNN models have two and three parallel VGG16 branches, respectively, which are concatenated before entering the classification network.

The FC-layers were tested with 512, 1024, 1536, 2048 and 4096 neurons, and dropout rates of 0, 0.3 and 0.5. This 15-model hyperparameter search was conducted on each of the three architectures, resulting in 45 models. These 45 models were run three consecutive times and averaged together for a more accurate result. Each model was trained using early stopping, stopping the model if validation loss did not decrease within 30 epochs. All model selections were based on the validation set performance. After training, the weight parameters from the epoch which performed best on the validation dataset were restored, and a final evaluation of the model was performed on the test dataset. The VGG16 networks had their weight parameters frozen during training. The model was written in Python 3.5 using the Keras machine learning library \cite{chollet2015keras}.

Table \ref{tab:result} shows the hyperparameters for the best performing models and their average result from the three consecutive runs. The MONO-CNN model achieves a result similar to that of the autoencoder. The two multiscale models perform equally and significantly better than the mono-scale models. The multiscale models also have a lower standard deviation on all metrics. Since both multiscale models achieve the same result, one could argue that the simplest model of the two should be chosen. In that case, DI-CNN with its 36M parameters is a simpler model than TRI-CNN which has 47M parameters in total. DI-CNN also have a marginally lower standard deviation than TRI-CNN.

\begin{table}[htbp]
	% The first argument is the label.
	% The caption goes in the second argument, and the table contents
	% go in the third argument.
	\floatconts
	{tab:result}
	{\caption{Models evaluated on the test set. F1-Score is presented as the total average and standard deviation calculated across all six classes over three consecutive runs. Parameters are shown as no. of trainable parameters / no. of total parameters.}}
	{\begin{tabular}{cccccc}
			Model & Input Scale & Dropout & FC-Neurons & \# Parameters & F1-Score \\ \hline
			Autoencoder\footnotemark\ & 400x 	& 0.1 & 256/512 & 89M/89M & 0.934 $\pm$ 0.061 \\
			MONO & 400x  						& 0.3 & 2048 & 5.3M/20M & 0.944 $\pm$ 0.007 \\
			DI & 100x/400x 						& 0.0 & 2048 & 6.3M/36M & 0.986 $\pm$ 0.002 \\
			TRI & 25x/100x/400x 				& 0.5 & 1024 & 2.6M/47M & 0.986 $\pm$ 0.003 \\ \hline
			
	\end{tabular}}
\end{table}

\footnotetext{Model trained and evaluated on the same dataset \cite{wetteland2019}.}

\section{Conclusion}
In this paper, we present preliminary results from a multiscale tile-based classification model. Tiles from six classes were extracted at multiple scales from WSI of patients diagnosed with urothelial carcinoma. Three model architectures were compared: MONO-CNN (400x), DI-CNN (100x, 400x) and TRI-CNN (25x, 100x, 400x). Results for an autoencoder model from previous work was also included for reference. Both multiscale models outperform the two single-scale models and achieve a very good result indicating the advantage of utilising multiple scales. The model can be used as an ROI extraction method for relevant tissue areas in the large WSI, useful for both pathologist and computer-aided diagnostic systems. Some more experiments should be performed to clarify if the behaviour stems from the multiscale approach or the extended field-of-view.

\bibliography{library}

\end{document}